\PassOptionsToPackage{prologue,table,xcdraw}{xcolor}
\documentclass[runningheads]{llncs}
\usepackage{times}

\usepackage{hyperref}

\usepackage[title]{appendix}
\usepackage{tablefootnote}
\usepackage{multirow}
\usepackage{amsmath}
\usepackage{siunitx}
\usepackage{amsfonts}
\usepackage{algorithmic}
\usepackage{algorithm}
\usepackage{diagbox}
\usepackage{booktabs}
\usepackage{makecell}
\usepackage{stackengine}

\renewcommand\cellgape{\Gape[1pt]}
\usepackage{textcomp}
\usepackage{xcolor}
\usepackage[numbers,sort&compress,sectionbib]{natbib}
\usepackage[nameinlink,capitalize]{cleveref}
\usepackage{hyperref}

\def\BibTeX{{\rm B\kern-.05em{\sc i\kern-.025em b}\kern-.08em
    T\kern-.1667em\lower.7ex\hbox{E}\kern-.125emX}}
    
\usepackage{graphicx} \usepackage[export]{adjustbox}
\usepackage{subfig}
\usepackage{epstopdf}
\DeclareGraphicsExtensions{.pdf,.png,.eps,.jpg,.jpeg,.tif,.tiff,.ps}
\graphicspath{{img//}}

\usepackage{soul}
\usepackage{xspace}
\definecolor{navy}{rgb}{0.1, 0.1, 0.8}
\definecolor[named]{gray}{rgb}{0.4, 0.4, 0.4}
\definecolor[named]{olive}{rgb}{0.1, 0.5, 0.1}
\definecolor[named]{ruby}{rgb}{0.8, 0.1, 0.3}
\definecolor{darkpastelgreen}{rgb}{0.01, 0.75, 0.24}
\definecolor{celestialblue}{rgb}{0.29, 0.59, 0.82}
\definecolor{coral}{rgb}{1.0, 0.5, 0.31}
\definecolor{Goldenrod}{rgb}{0.8,0.8,0}

\newcommand{\eat}[1]{}

 \newcommand{\NOTE}[2]{}
 \newcommand{\TODO}[2]{}
 \newcommand{\nb}[1]{}
 \newcommand{\mar}[1]{}
 \newcommand{\cl}[1]{}

\newcommand{\squishlist}{
 \begin{list}{$\bullet$}
  { \setlength{\itemsep}{0pt}
     \setlength{\parsep}{3pt}
     \setlength{\topsep}{3pt}
     \setlength{\partopsep}{0pt}
     \setlength{\leftmargin}{1.5em}
     \setlength{\labelwidth}{1em}
     \setlength{\labelsep}{0.5em} } }

\newcommand{\squishlisttwo}{
 \begin{list}{$\bullet$}
  { \setlength{\itemsep}{0pt}
    \setlength{\parsep}{0pt}
    \setlength{\topsep}{0pt}
    \setlength{\partopsep}{0pt}
    \setlength{\leftmargin}{1.5em}
    \setlength{\labelwidth}{1.5em}
    \setlength{\labelsep}{0.5em} } }

\newcommand{\squishend}{
  \end{list}  }

\newcommand{\titlename}{Linking the Dynamics of User Stance \\ to the Structure of Online Discussions}

\begin{document}
\title{\titlename}
\author{Christine Largeron\inst{1} \and
Andrei Mardale\inst{1} \and \\
Marian-Andrei Rizoiu\inst{2}$^{,\href{https://orcid.org/0000-0003-0381-669X}{0000-0003-0381-669X},*}$}
\authorrunning{C. Largeron et al.}

\institute{Univ Lyon, UJM-Saint-Etienne, CNRS, OGS, France
\email{Christine.Largeron@univ-st-etienne.fr} 
\and
Data Science Institute, University of Technology Sydney, Australia \\
\email{Marian-Andrei.Rizoiu@uts.edu.au}; $^*$corresponding author
}
\maketitle              \begin{abstract}
This paper studies the dynamics of opinion formation and polarization in social media. We investigate whether users' stance concerning contentious subjects is influenced by the online discussions they are exposed to and interactions with users supporting different stances. We set up a series of predictive exercises based on machine learning models. Users are described using several posting activities features capturing their overall activity levels, posting success, the reactions their posts attract from users of different stances, and the types of discussions in which they engage. Given the user description at present, the purpose is to predict their stance in the future. Using a dataset of Brexit discussions on the Reddit platform, we show that the activity features regularly outperform the textual baseline, confirming the link between exposure to discussion and opinion. We find that the most informative features relate to the stance composition of the discussion in which users prefer to engage.

\end{abstract}
\nocite{Mishra2018}
\section{Introduction}
\label{sec:Intro}

\addtocounter{footnote}{-1}\footnotetext{\label{fn:supp-material}Supplementary Information available online: \url{https://arxiv.org/pdf/2101.09852.pdf\#page=13}}
\addtocounter{footnote}{1}\footnotetext{\label{fn:code-data}Code and data publicly available: \url{https://github.com/behavioral-ds/online-opinion-dynamics}}

In the twenty-first century, offline events are increasingly shaped by the discussions occurring on online social media.
The outcome of significant events --- such as the presidential elections in the United States of America~\cite{hughes2009twitter,Rizoiu2018a,Kim2019} or the decision of the United Kingdom to leave the European Union~\cite{howard2016bots} --- were influenced by the opinions that voters formed using a wide array of online sources, including on social media.

Contentious subjects usually lead to heated arguments on social media, which in turn polarize public opinion.
The prevailing theory is that online polarization emerges due to filter bubbles, which only expose users to peers with the same views~\cite{Banisch2019}.
This led to a body of work that believes that online polarization can be addressed by exposing users to contrary news and views~\cite{Garimella2017a,Graells-Garrido2013,Matakos2017}. 
However, participatory studies concluded that exposure to opposing views on social media could increase polarization~\cite{Bail2018,Liao2013}.
There is still an open gap concerning opinion formation on social media.

This work addresses two specific open questions concerning the dynamics of polarized opinion formation in the context of Reddit discussions around Brexit.
The first open question deals with how users form polarized opinions.
Some works claim that social media increases polarization \cite{Messing2014,De-Wit2019}, while other studies find that the usage of social media reduces polarization~\cite{Barbera2014}.
Furthermore, participatory and measurement studies~\cite{Dubois2018} challenge this idea altogether, indicating that information savvy people leverage diverse sources of information and escape the filter bubble.
The question is \textbf{are the stances of users concerning contentious subjects influenced by the discussions they are exposed to?}
The second open question focuses on the dynamics of polarization.
Previous work concentrates mainly on detecting and forecasting opinion polarization based on content diffusion in online social networks~\citep{Garimella2016,Rama2017}; little work concentrates on detecting polarization dynamics.
\textbf{Can we predict the future stance of users based on their present activity? and what are the factors that influence the changes of stances?}

We answer the questions mentioned above on a longitudinal dataset containing discussions around Brexit on Reddit, spanning from November 2015 until April 2019.
Our work assumes two factors that determine users' stance towards contentious subjects.
First, user stance has inertia, i.e., the stance at a given time is dependent on their past stance.
Second, user stance depends on the stance of other users with whom the said user interacts.
Consequently, the interactions with users of known stances indicate the future user stance, even without observing the textual content of these interactions.

We first divide the dataset time extent into fourteen time-periods, based on the notable events in the real-life Brexit timeline, such as the referendum, the triggering of Article 50, or the EU rejecting the UK's white paper.
We investigate whether users' stance concerning contentious subjects is influenced by the online discussions they are exposed to and interactions with users supporting different stances. 
As there are no annotations available, we transfer a textual classifier trained on Twitter data to classify user stances in Reddit.
Next, we answer the first open question by building three feature sets to describe user activity during each period.
The purpose of these features is to capture a user's interaction with the other users of known stance in the community.
The constructed features include overall activity levels, posting success, the reactions their posts attract from users of different stances, and the types of discussions in which users engage.
We answer the second open question by setting up a series of predictive exercises that forecast the user stance in the next period based on the user description in the current period. 
We show that the activity features regularly outperform the textual baseline, indicating that user opinions are influenced by the discussions they are exposed to. 
We find that the discussion's stance composition that users prefer to engage in is the most informative feature.
Notably, the content posted by a user during a time period appears to be less informative about the next period's user stance.

\textbf{The main contributions of this work are as follows:}
\squishlist
	\item We propose three feature sets predictive of the user stance that leverage solely the structure of the discussion (i.e., not the textual content emitted by the user).
	\item We show that all three feature sets are more predictive of the future stance than a textual baseline trained on the content emitted in the present.
\item We provide predictive evidence that user polarization dynamics are linked to the stance composition of the discussions that the users are exposed to. 
\squishend

\section{Related Work}
\label{sec:State}

We structure the discussion of the related works into two categories:
detecting and alleviating polarization,  
and opinion and polarization dynamics.

\noindent\textbf{Detecting and alleviating online polarization.}
Previous work concentrates on detecting and reducing online polarization.
Detection methods usually start from the social graph of the users.
If the graph is presented as a signed network --- i.e., the nodes are users, and the edges between users of the same polarity have a positive sign, while the edges across two polarities have a negative sign --- community detection uncovers polarized communities~\citep{Bonchi2019}.
The idea is to search for two communities (subsets of the network vertices) where within communities there are mostly positive edges while across communities there are mostly negative edges.
When the sign of edges is not available, \citet{Garimella2017a} propose to use the diffusion cascades that occur on top of the social graph to detect the communities of users that participate together in the same cascades.
Finally, they create a controversy score for discussion topics based on how polarized apart the communities are.
In this work, we use a supervised approach to detect user polarization based on their emitted text:
we use a textual classifier trained on annotated Twitter data to label the stance of Reddit users.

When it comes to reducing online polarization, it is generally assumed that exposing users to opposite views reduces their polarization~\cite{Matakos2017,Gillani2018}.
\citet{Garimella2017} devised tools and algorithms to bridge the polarized echo chambers and reduce controversy. 
They represent online discussions on controversial issues with an endorsement graph, and they cast the problem as an edge-recommendation problem on this graph.
\citet{Graells-Garrido2013} study how to take advantage of partial homophily to suggest agreeable content to users authored by people with opposite views on sensitive issues, while \citet{Musco2018} search for the structure of a social network that minimizes disagreement and controversy simultaneously.
However, empirical studies appear to contradict the fundamental thesis that users exposed to contrary views temper their polarization.
\citet{Bail2018} performed a participatory study on Twitter users, where they paid users to follow bots emitting tweets of the opposing opinion.
They found that most users reinforced their previously held opinions and that exposure to opposing views on social media can increase political polarization.

\noindent\textbf{Opinion and polarization dynamics.}
The prior work most relevant to this paper concerns the political polarization around Brexit and the study of polarization dynamics.
\citet{Grcar2017} studied the relation between the Twitter mood and the referendum outcome and who were the most influential Twitter users in the Pro- and Against- Brexit camps. 
They constructed a stance classification model, and they predicted the stance of about one million UK-based Twitter users. 
They found that the top pro-Brexit communities are considerably more polarized than the contra-Brexit camp.
\citet{benoit_network_nodate} collected 23 million Tweets related to the EU referendum in the UK to predict the Brexit vote. 
They used user-generated hashtags to build training sets related to the Leave/Remain campaign, and they trained an SVM to classify tweets.
The above work uses textual content to decide the stance of a user.
In contrast, our work leverages the structure of the discussion in which users engage without observing the textual content.
In our experiments in \cref{sec:results} we show that our methods consistently outperform content-based methods.

When modeling the dynamics of opinion polarization, \citet{Das2014} start from the conformity theory -- i.e., a user will adopt the majority of their neighbors' opinion -- and propose a biased voter model. 
They show preliminary theoretical and simulation results on the convergence and structure of opinions in the entire network.
On the empirical side, longitudinal study of controversy on Twitter~\cite{Garimella2017c} did not find long-term trends.
However, they find that for particular subjects, polarization increased.
By comparison, our work deals with the short-term polarization dynamics: we are interested in how users update their polarity concerning controversial topics based on their exposure to the content of different polarities.

\section{The Dynamics of User Stance and Dataset}
\label{sec:Notation_dataset}

This section introduces our hypotheses around the dynamics of user stances, the structure of online discussions (on Reddit), and the dataset that we collected for the Brexit case study. 

\noindent\textbf{User stance dynamics.}
When faced with contentious subjects, users usually have opinions --- dubbed here as \emph{stances}; in the case of our study (i.e., Brexit) we define the following set of stances:
\textbf{A}gainst-Brexit, \textbf{P}ro-Brexit, or \textbf{N}eutral.
Our work's central hypothesis is that users can update their stance as time passes by (for example, from Neutral to pro- or against- Brexit).
Furthermore, we hypothesize that the change occurs partly due to the discussions the users are exposed to at present.
We posit that stance changes occur on a much longer time scale than that of diffusions and threads.
Without loss of generality, we assume that the time extent $\mathcal{O}$ of our dataset is divided into periods (or time frames) during which each user's stance is constant. 
The time periods are defined by a set of cutoff times $o_j \in \mathcal{O}$ such that $[o_j, o_{j+1}]$ defines an interval t. 
We denote the stance of a user $u$ in a given time interval $t$ as $c_t(u) \in \{A, P, N \}$.
When passing from one period to the next, the users update their stance or maintain the stance from the previous period -- in other words, the user $u$ updates their stance from $c_t(u)$ to $c_{t+1}(u)$.

\begin{figure*}[t]
	\centering
	\includegraphics[width=.99\textwidth]{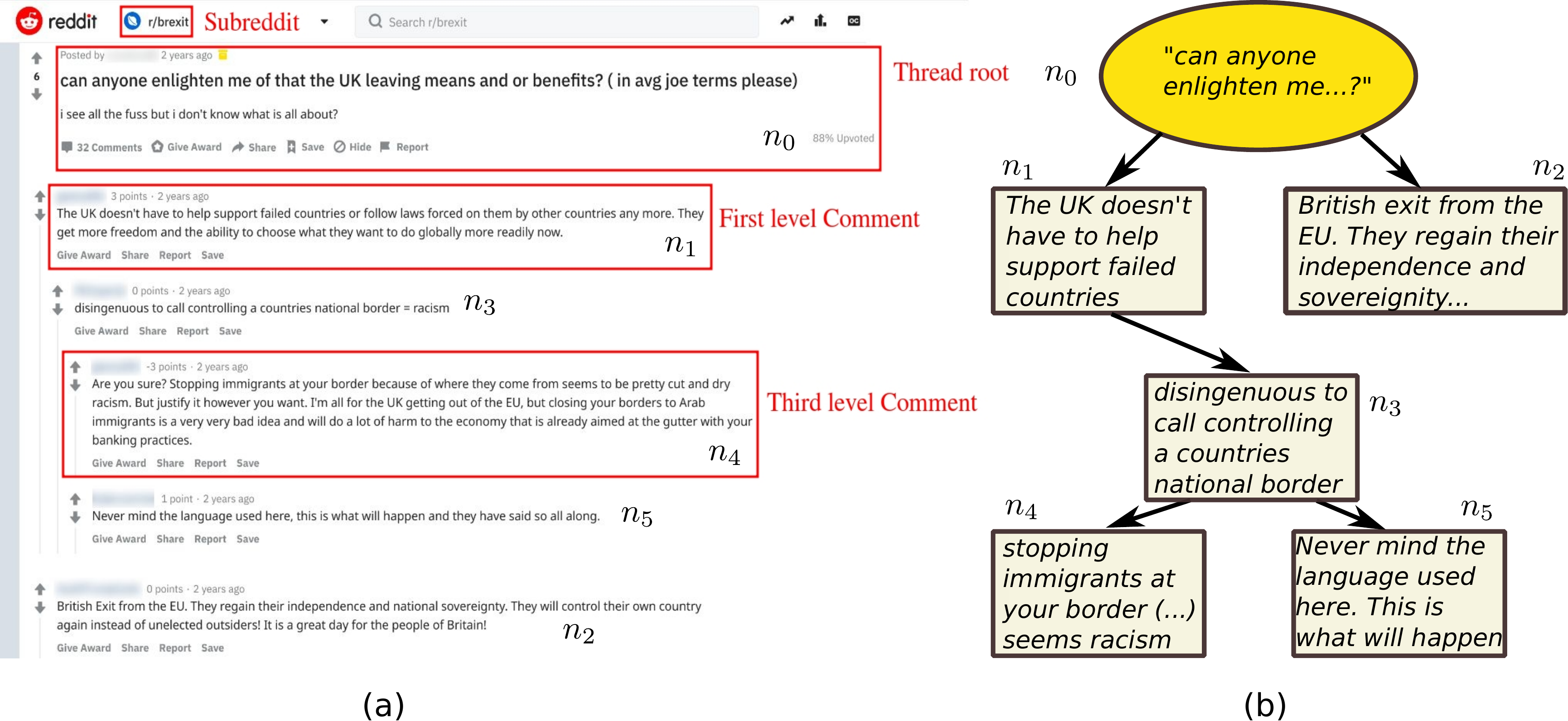}\hfill
	\caption{ 
		\textbf{(a)} Elements of the Reddit platform. Structure of a discussion thread, with multi-level comments, inside a subreddit. 
		\textbf{(b)} Logical structure used for analyzing the data.}
	\label{fig:example-logical-structure}
\end{figure*}

\noindent\textbf{Structure of discussion on online social media.}
Online social networks can be viewed as meeting places where users have online discussions, submit content and articles in the form of text, link or media.
In these meeting places, users interact with their peers, form and update opinions and stances towards topics.
For example, on Reddit, users can start threads similar to forum environments or post comments on existing threads.
Consequently, the discussions present themselves as hierarchies of posts in a tree-like structure.
\cref{fig:example-logical-structure}a shows an example of a real Reddit discussion, containing an initial post ($n_0$) and five comments ($n_1$ to $n_5$).
For instance, comment $n_3$ is a reply to comment $n_1$.
The resulted tree structure is shown in \cref{fig:example-logical-structure}b, and leveraged in \cref{subsec:predictive-features} to construct the non-textual features describing the activity of users.
In the rest of this paper, we denote a tree of posts as \emph{a thread}, which is started by \emph{a post} -- also known as \emph{submission} in Reddit terminology, the root of the tree.
We denote all the other nodes as \emph{comments} -- 
chronologically subsequent messages posted as replies to a post or other comments in the same thread.

We collectively denote posts and comments as \emph{entries}.
An entry is a triplet $s_j = (u_j, pc_j, d_j)$, where $u_j$ denotes the user name, $pc_j$ the published content, and $d_j$ is the time stamp of the entry $s_j$.
We further define a \emph{diffusion} $\delta_i$ as a temporally ordered sequence of entries, starting with a post, ending with a leaf comment and containing all comments on the path connecting the post and the leaf comment. 
Formally, it is defined as $\delta_i = \{ s_j = (u_j, pc_j, d_j) | j = 1, 2, .. \} $.
Visibly, there are as many diffusions in a thread as there are leaf nodes.
For example, in \cref{fig:example-logical-structure}b there are three diffusions: $\{n_0, n_1, n_3, n_4\}$, $\{n_0, n_1, n_3, n_5\}$ and $\{n_0, n_2\}$.
Finally, a thread is a set of diffusions $\mathcal{S} = \{\delta_i | i = 1, .., N\}$.

\noindent\textbf{Dataset: Brexit discussions on Reddit.}
We collected the Reddit dataset used in our case study using the Pushshift API \cite{pushshift}. 
It contains 229,619 entries (21,725 posts and 207,894 comments) posted between November 2015 and April 2019 on the \emph{brexit} subreddit (\url{https://www.reddit.com/r/brexit/}). 
Each entry has the following variables:
entry id, text, timestamp, author, parent id (useful for building the tree structure as shown in \cref{fig:example-logical-structure}a), Reddit score, and the number of comments for the entry.
A total of 14,362 unique authors participated in these discussions.
We have divided the dataset's time extent into 15 intervals based on the occurrence date of real events, such as the UK referendum of 23 June 2016, the nomination of M. Barnier as Chief Negotiator, beginning of the Brexit negotiations, rejection of the UK white paper by EU, the publication of the Brexit withdrawal agreement, first and second meaningful votes, etc. 
We split the entries into 15 subsets according to the time interval in which they were posted.
Due to space constraints, we further profile the dataset and the 15 intervals in the online supplement$^{\ref{fn:supp-material}}$.
Also note that the constructed dataset, together with the code to build the feature sets detailed in \cref{subsec:features} are publicly available$^{\ref{fn:code-data}}$.

\section{Forecast User Stance Dynamics}
\label{subsec:predictive-features}

This section tackles the two research questions by posing them as supervised machine learning problems.
We first describe the learning problem (\cref{subsec:machine-learning});
next, we build predictive features that embed user interactions with users of different stances (\cref{subsec:features});
finally, we describe the predictive setup (\cref{subsec:predictive-setup}).

\subsection{A Supervised Machine Learning Problem}
\label{subsec:machine-learning}
We cast the problem of forecasting the future stance of users as a supervised machine learning problem.
Each user $u$ is represented by a set of features $FS_t(u)$ describing her Reddit activity during the time interval $t$.
The feature set also includes the user stance at the current time $t$, i.e., $c_t(u)$.
The task consists in forecasting the user stance at the next time interval $t+1$, i.e., $c_{t+1}(u)$, using the features at time $t$, i.e., $FS_t(u)$.
Off-the-shelf classifiers are used to learn a model from $FS_t(u)$ to $c_{t+1}(u)$.
The difficulty lies in defining the features that describe the user's activity during a period and obtaining the ground truth labels to build the training set and the test set. 
To determine user stances, we use a textual classifier trained on Twitter data (further detailed in \cref{subsec:twitter-stance-predictor}).
In the next section, we design several feature sets that capture users' activity and their interactions with other users of different stances. 

\begin{table}[tbp]
	\small
	\newcommand\widthcellone{4cm} \newcommand\widthcelltwo{8cm} \renewcommand\cellgape{\Gape[2pt]}
	\centering
	\caption{Constructed feature sets describing user interactions with information diffusions.}
	\label{tab:constructed-features}
	\begin{tabular}{p{\widthcellone}p{\widthcelltwo}}
		\toprule
		\textbf{Feature set} & \textbf{Features} \\ \midrule
		\makecell[t{p{\widthcellone}}]{\textbf{FS1}\\(User activity)} & \makecell[t{p{\widthcelltwo}}]{
			-- number of initiated diffusions $ID_t(u)$ \\
			-- number of submitted comments $CS_t(u)$\\
			-- quantiles of the number of received comments per entry $R_t^1(u),...,R_t^5(u)$\\
			-- stance at current time-frame $c_t(u)$.} \\ 
			
	\makecell[t{p{\widthcellone}}]{\textbf{FS2}\\(User activity per stance)} & \makecell[t{p{\widthcelltwo}}]{
			-- number of comments submitted to entries from each stance $CS_t^A(u), CS_t^P(u), CS_t^N(u)$\\
			-- quantiles of the number of received comments per entry, tallied by commentator stance $R_t^{x1}(u),..., R_t^{x5}(u)$,  $x \in \{A, P, N\}$\\
			-- stance at current time-frame $c_t(u)$\\
			}\\ 
			
		\makecell[t{p{\widthcellone}}]{\textbf{FS3}\\(Structure of diffusion)} & \makecell[t{p{\widthcelltwo}}]{
			-- quantiles of the number of submitted comments in diffusions per stance \	$UP_t^{x1}(u), ..., UP_t^{x5}(u)$, $x \in \{A, P, N\}$\\
			-- stance at current time-frame $c_t(u)$}\\

		\textbf{FS4} (Relational features) & FS1 + FS2 + FS3 \\
	
		\textbf{FS0} (Textual features) & 100 top words + $c_t(u)$ \\
		
		\textbf{FS5} (All features) & FS0 + FS4 \\
		
		\bottomrule
\end{tabular}
\end{table} 
\subsection{Predictive Features}
\label{subsec:features}
We introduce three sets of features (denoted as \textbf{FS1}, .., \textbf{FS4}, shown in \cref{tab:constructed-features}) aimed at capturing increasingly complex information concerning user activity.
\textbf{FS1} serves as an activity baseline, tallying user posting activity and the comments they receive.
\textbf{FS2} aims to capture how the user interacts with users of different stances (e.g. do they prefer to comment on entries with similar stances to their own? to the opposite stance?), and whether they elicit more comments from users with the same polarity or the opposite.
\textbf{FS3} aims to capture the type of threads in which the user engages (e.g., do they like to engage in discussion with a single stance or deliberative threads?).
We detail each set.

\textbf{FS1} focuses on the activity of the user at the global level. 
For a given user $u$ and a time interval $t$, we count $ID_t(u)$, the number of diffusions initiated by $u$ during the interval $t$ (\textit{i.e.}, the number of posts sent by $u$) and $CS_t(u)$, the number of comments submitted by $u$ during the interval $t$ by excluding auto-comments. Thus, the number of entries submitted by $u$ during the period is denoted $N_t(u)$ with $N_t(u) = ID_t(u) + CS_t(u)$.
We also consider the user's success defined as the number of replies generated by his activity and quantified by the direct or indirect comments received by each entry (post or comment) submitted by $u$ during the period. 
Formally, if $r_i$ denotes the number of replies following the entry $m_i$ submitted by $u$ during the period $t$, we obtain the set $\{ r_i | i = 1,.., N_t\}$ and we compute the quantiles $R_t^1(u), ..., R_t^5(u)$ corresponding respectively to $0\%$, $25\%$, $50\%$, $75\%$ and $100\%$ of its distribution. 
Thus, \textbf{FS1} contains 8 features including $c_t(u)$ (the user stance at the current time).

\textbf{FS2} aims to capture how the user interacts with users of different stances: \textbf{A}gainst, \textbf{P}ro or \textbf{N}eutral in our case study.
First, we measure how the user engages with content from other users by counting the comments sent by user $u$ during the period $t$ in response to entries posted by each group denoted respectively $CS_t^{A}(u), CS_t^{P}(u)$ and $CS_t^{N}(u)$. 
Thus, $CS_t(u) = CS_t^{A}(u) + CS_t^{P}(u) + CS_t^{N}(u)$, where $CS_t(u)$ has been defined in \textbf{FS1}.
The underlying idea is to capture whether $u$ exchanges more with users having the same stance as him or with users having a different stance.
Second, we measure how the users of the different stances engage with $u$ by counting the number of comments received from each group in response to entries sent by $u$ during the period $t$. 
Thus, if $r_i^{x}$ denotes the number of replies from group $x \in \{A, P, N\}$ following the entry $m_i$ submitted by $u$ during the period $t$, we obtain the distribution $\{ r_i^{x}, i = 1,.., N_t\}$ and we compute the quantiles $R_t^{x1}(u), ..., R_t^{x5}(u)$ corresponding respectively to $0\%$, $25\%$, $50\%$, $75\%$ and $100\%$ of this distribution.
With this second set composed of 19 features, the objective is to capture whether content emitted by $u$ attracts comments from the group of users of similar stance or from the other stances.

\textbf{FS3} aims to capture the type of threads in which the user $u$ engages.
For each threads in which $u$ posted an entry (post or comment) during the period, we compute the number of entries per group. More precisely, if $NS$ denotes the number of threads in which $u$ sent at least one entry during the period and $S_i$ is one of these threads, we compute the number of entries $A_i, P_i, N_i$ respectively emitted by each group in $S_i$. By this way, we obtain three sets $\{A_i | i = 1, .., NS\}, \{P_i | i = 1, .., NS\}, \{N_i | i = 1, .., NS\}$ that we can summarized by their respective quantiles $UP_t^{x1}(u), ..., UP_t^{x5}(u)$, $x \in \{A, P, N\}$. 
Thus, if a user with a given stance, for example \textbf{A}nti-Brexit, prefers to exchange with the other anti-Brexit users, the features $UP_t^{A1}(u)$, ..., $UP_t^{A5}(u)$ will have higher values than $UP_t^{P1}(u)$, ..., $UP_t^{P5}(u)$, $UP_t^{N1}(u)$, ..., $UP_t^{N5}(u)$.
So, \textbf{FS3} contains 16 features, including $c_t(u)$.
We also build \textbf{FS4}, the union of the above mentioned feature sets: $\textbf{FS4} = \textbf{FS1} \cup \textbf{FS2} \cup \textbf{FS3}$.

We also build a textual baseline based on the user's content in the current period. 
We first extract the top 100 most frequent words (stop words removed) from all the Reddit dataset entries over all the time intervals.
Next, we aggregate the text of all the entries of each user into a single document, and we compute the TF-IDF scores for the selected top 100 most frequent words.
Consequently, \textbf{FS0} contains 101 features, including the user stance at the current time-frame $c_t(u)$.
Finally, we also consider \textbf{FS5} composed of all the textual and relational features: $\textbf{FS5} = \textbf{FS4} \cup \textbf{FS0}$.

\subsection{Learning Stance in Twitter} 
\label{subsec:twitter-stance-predictor}

One of the main challenges of this work is the lack of ground truth, \textit{i.e.}, the stance for Reddit users at each time interval. 
We transfer to our Reddit dataset a model trained on Twitter and initially introduced by \citet{benoit_network_nodate}.

\noindent\textbf{The Twitter dataset.}
\citet{benoit_network_nodate} collected the Twitter dataset from 6 January 2016 to July 2016 using the Twitter Firehose API.
They crawled all the tweets using three search criteria related to Brexit: the general search term \textit{Brexit}, hashtags such as \textit{\#leaveeu} or \textit{\#yes2eu} and Twitter usernames of groups and users set up to communicate about Brexit (e.g., \textit{@voteleave} or \textit{@yesforeurope}. 
The resulted dataset contains 26.5 million tweets emitted by 1.5 million users.

\noindent\textbf{Build a stance predictor in Twitter.}
We build a Twitter stance predictor following the methodology proposed by \citet{benoit_network_nodate}, which we briefly summarize below.
\citet{benoit_network_nodate} currated two sets of hashtags, shown in \cref{tab:table_hashtags}, which indicate the user stance and utilized in 136 thousand tweets.
\begin{table}[tbp]
	\centering
	\newcommand\mywidthleft{1cm} \newcommand\mywidthright{11cm} 

	\caption{Hashtags used by \citet{benoit_network_nodate} for splitting Twitter users in two categories, to train the Naive Bayes Classifier.}
	\label{tab:table_hashtags}

	\begin{tabular}{p{\mywidthleft}p{\mywidthright}}
		\toprule
		\textbf{Stance} & \multicolumn{1}{c}{\textbf{Hashtags}} \\ \midrule
		\makecell[t{p{\mywidthleft}}]{\textit{Pro Brexit}} & 
		\makecell[t{p{\mywidthright}}]{
			\#voteleave \#inorout \#voteout \#takecontrol \#borisjohnson \#lexit \#independenceday \#ivotedleave \#projectfear \#britain \#boris \#go \#projecthope \#takebackcontrol \#labourleave \#no2eu \#betteroffout \#june23 \#democracy
		} \\ 
		\makecell[t{p{\mywidthleft}}]{\textit{Against Brexit}} & 
		\makecell[t{p{\mywidthright}}]{
			\#strongerin \#intogether \#infor \#votein \#libdems \#voting \#incrowd \#bremain \#greenerin
		} \\ 
\bottomrule
\end{tabular}
\end{table} We filter out occasional users -- who emit less than 50 tweets -- and users who do not employ any of the hashtags.
For each of the remaining 11,277 users, we compute a `leave' score equal to the difference between the number of used Pro-Brexit hashtags and Against-Brexit hashtags.
We rank the users based on the score, and we select the 10\% users with the lowest (negative) score as Against-Brexit users and the 10\% of users with the highest (positive) score as Pro-Brexit users.
The resulting set contains the aggregated tweets (one document per user) for 2,257 users.
We first perform the usual text preprocessing: we remove stopwords, punctuation signs, hashtags, mentions, and other diacritics; we convert all letters to lower case, remove rare words, and perform stemming.
Next, we train a Naive Bayes classifier using $80\%$ of the data, and we evaluate using the remaining $20\%$ of the data. 
The model outputs the probability for a document (\textit{i.e.}, a user) to belong to one of the classes (Against-Brexit or Pro-Brexit). 
Following the methodology proposed by \citet{benoit_network_nodate}, we convert this output probability into a discrete label: 
if the \emph{leave} probability is below $0.25$, we label the user as Against-Brexit; 
if it is greater than $0.75$, we label the users as Pro-Brexit. 
Otherwise, the label is Neutral. 
On the test set, the trained model obtains a prediction macro-accuracy of $89.36\%$ and a macro-F1-score of $88.68\%$. 
As shown in the next section, we transfer the trained model to compute $c_t(u)$, the users' stance in each period in the Reddit dataset.
We use a Naive Bayes classifier because it is somewhat robust to concept drift and noisy features~\cite{schutze2008introduction} -- here, vocabulary change between Twitter and Reddit.
The robustness is because rank scores are typically correct even if the conditional independence assumption is violated. 
We use cut-offs on the Naive Bayes score rather than interpreting the score as a probability in absolute terms.

\subsection{Predictive Setup}
\label{subsec:predictive-setup}

\textbf{Building Reddit learning and testing sets.}
For each timeframe, we first aggregate all the Reddit messages of each user into a single document.
Next, we assign them a Brexit stance using the Naive Bayes classifier trained on the Twitter dataset (detailed in \cref{subsec:twitter-stance-predictor}).
As we perform this procedure for each interval, we obtain not only the present stance of the user $c_t(u)$ but also the stance at the next timeframe $c_{t+1}(u)$. 
Finally, we compute the predictive feature sets $\textbf{FS1},...\textbf{FS5}$ for each user and each period from the Reddit dataset.

\noindent\textbf{Models and evaluation.}
We predict users' stance in the next timeframe using each feature set computed on the current timeframe.
We train and test five different algorithms -- Logistic Regression, KNN, Random Forest, Gradient Boosting~\cite{friedman2001greedy}, XGBoost~\cite{chen2016xgboost}.
We evaluate using a double Cross-Validation. 
First, we use a 10-fold outer Cross-Validation to split the data into training and testing sets. 
At each fold, we tune hyper-parameters using an inner 5-fold Cross-Validation together with Random Search with 500 iterations.   
We measure performance using standard evaluation metrics and their standard deviation:
macro-F1, macro-Accuracy, macro-Precision, and macro-Recall.

\section{Results}
\label{sec:results}

\begin{figure*}[tbp]
	\centering
	\newcommand\myheightA{0.18} \subfloat[]{
		\includegraphics[height = \myheightA\textheight]{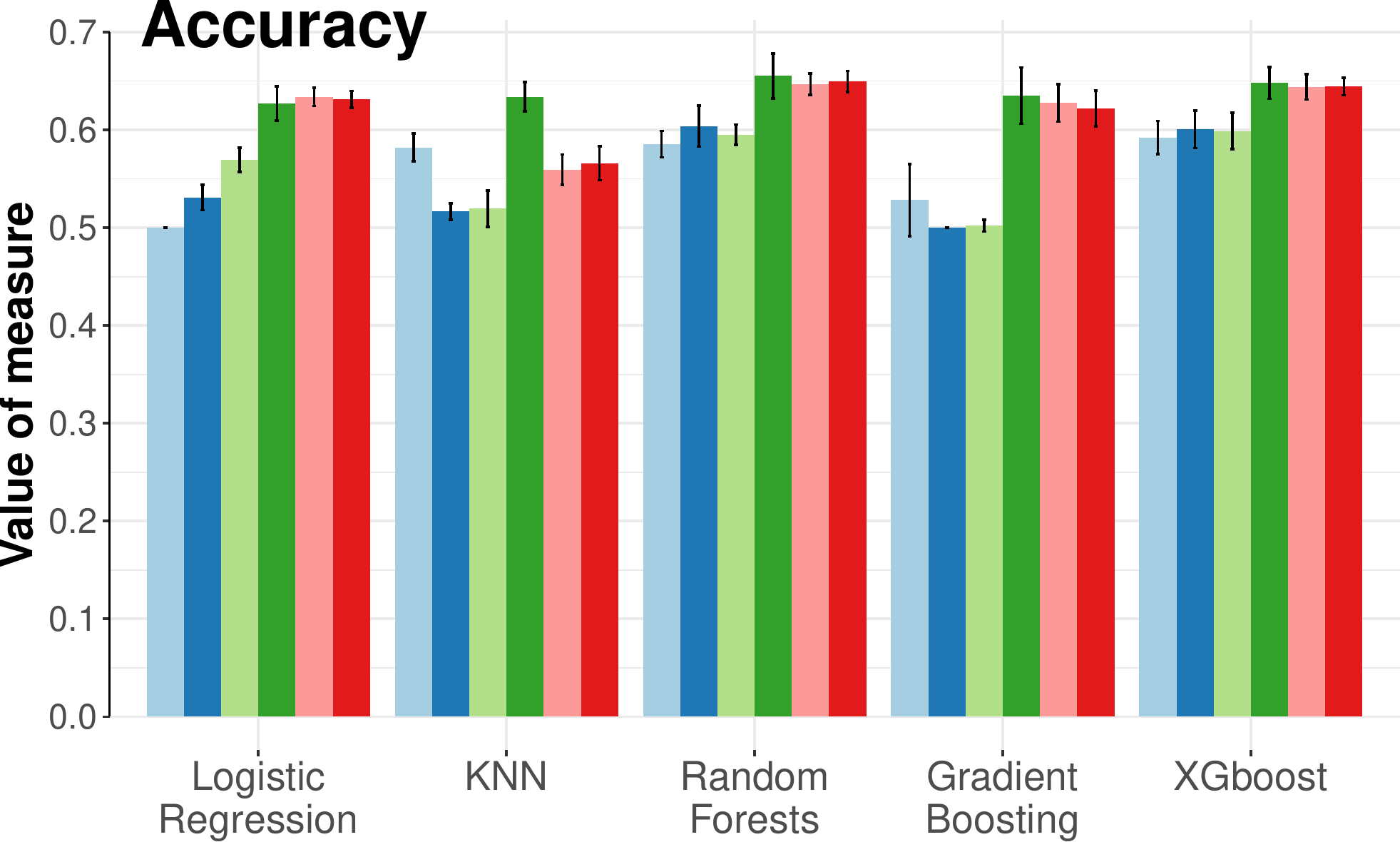}
\label{subfig:accuracy}}\subfloat[]{
		\includegraphics[height = \myheightA\textheight]{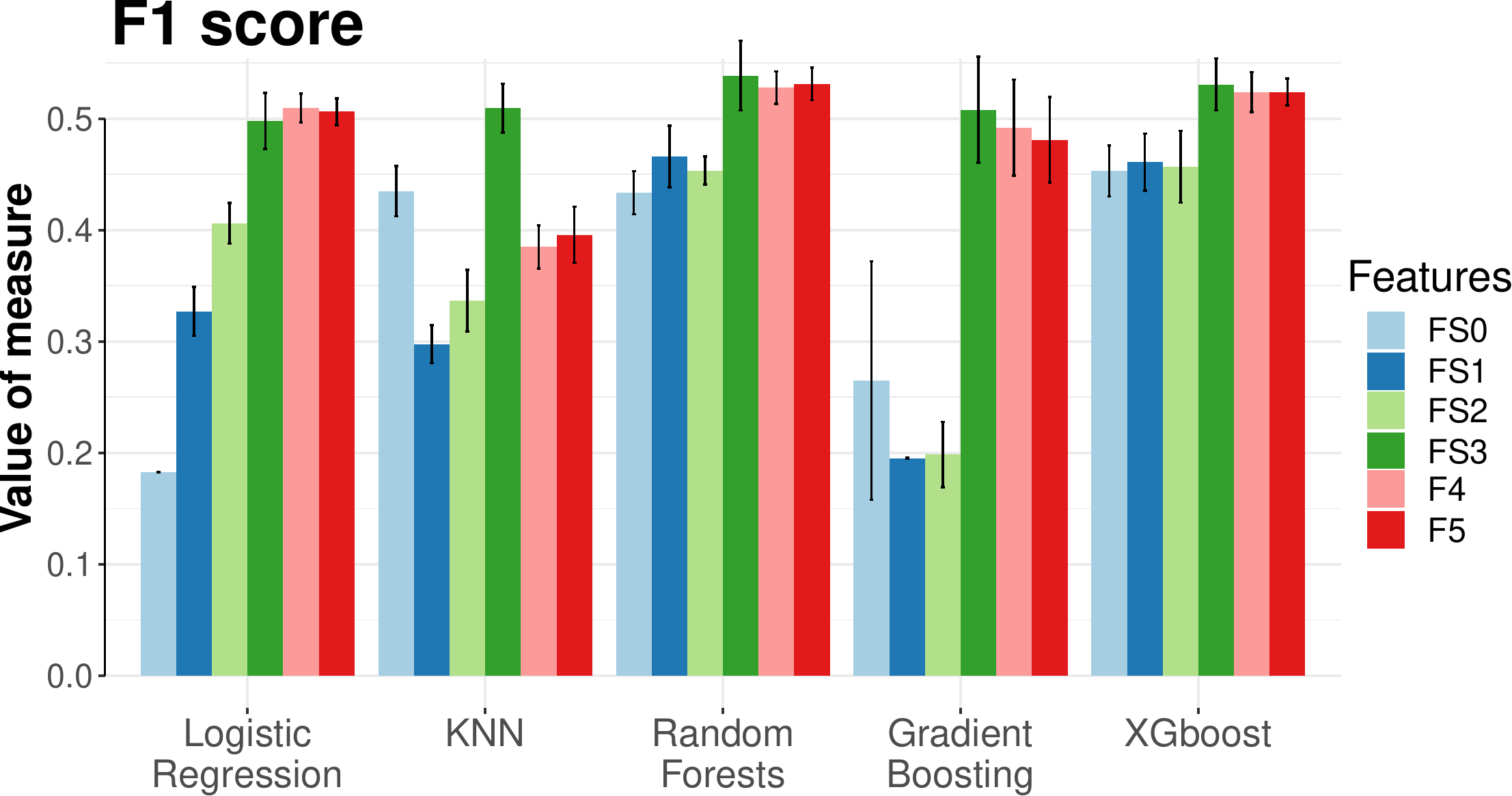}\label{subfig:F1}}\caption{
		Evaluation metrics for the developed models: accuracy \textbf{(a)} and F1-Score \textbf{(b)}.
	}
	\label{fig:figure_11}
\end{figure*}

This section presents the obtained performances for predicting the future stance of users.
The Reddit dataset is imbalanced, with most of the user having a Neutral stance.
Therefore, \cref{fig:figure_11} plots the macro versions of accuracy and F1 score (macro-precision and macro-recall are shown in the Supplementary Information$^{\ref{fn:supp-material}}$).
Note that we use the macro version of the metrics, which gives equal representation to minority classes and alleviate the class imbalance in our dataset.
Note that for a three-class classification problem (here Against, Neutral, Brexit), an unweighted, random classifier is expected to obtain an F1 score and accuracy score of $33\%$.

\textbf{Analysis of the relational features.}
\cref{fig:figure_11} shows that the best classifier reaches $53.9\%$ F1 score, which is double the random score. 
As the data is imbalanced, the accuracy is higher at $65\%$.
For most classifiers, the performance improves from FS1 to \textbf{FS3}, with \textbf{FS3} providing the best performance for all methods, except KNN.
This indicates that the stance composition of the threads that the user prefers to engage in best indicates her future stance.
The best performing classifiers are Random Forest and XGBoost.
Interestingly, the combination of all activity features (denoted as \textbf{FS4}) does not further improve results.

\textbf{Relational and textual features.}
\cref{fig:figure_11} shows that relational features (\textbf{FS1} to \textbf{FS4}) have higher predictive power than textual features (FS0), for the best performing method (Random Forest and XGBoost).
While the conclusions are more nuanced for the other classifiers, \textbf{FS3} outperforms FS0 for all classifiers and all metrics.
This result suggests that the type of discussions users engage in indicates their future stance more than the content they emit at present.
Moreover, we observe that using textual together with relational features (\textbf{FS5}) does not improve results significantly as the performances of \textbf{FS5} are equal to \textbf{FS3}. 

\textbf{Analysis per stance.}
We analyze in more detail the performances of the best performing classifier (XGBoost) on the best features set (\textbf{FS3}).
We compute the prediction performances for each combination of present and future stance -- i.e., the nine combination $ \{ (c_t(u), c_{t+1}(u)) | c_t(u), c_{t+1}(u) \in \{A, B, N\} \} $.
The values are reported in \cref{fig:table_f1_scores}. 
We see that the classifier performs well for the users who maintain their opinion between two subsequent timeframes (shown by the main diagonal of \cref{fig:table_f1_scores}). 
Noteworthy, it also performs well for the transitions from Neutral to Pro- or Against-Brexit, with F1 scores equal to $0.51$ and $0.45$, respectively. 
The result implies that we can predict the future stance of the currently undecided participants in online debates.
The implications are significant, as most democratic processes tend to be decided by swaying undecided voters.

\begin{table}[tbp]
	\centering
	\caption{F1 score of predicting next stance, tabulated per current stance.}
\label{fig:table_f1_scores}
	\begin{tabular}{l|ccc}
		\toprule
		\backslashbox{\textit{Stance at t}}{\textit{Stance at t+1}}
		&\textbf{Against} & \textbf{Neutral} & \textbf{Brexit}\\ \midrule
		\textbf{Against} &0.68&0.34&0.35\\
		\textbf{Neutral} &0.51&0.62&0.45\\
		\textbf{Brexit} &0.44&0.34&0.59\\ \bottomrule
\end{tabular}
\end{table}

\section{Conclusion}
\label{sec:conclusion}

In this paper, we analyzed information diffusion in social media platforms, and we studied whether the stances of users are influenced by the discussions to which they are exposed.
To capture the dynamics of the opinions of online communities, we chose the Reddit platform and Brexit as a case study due to its polarity. 
To better understand why users change their stance, we predict the future user stance using supervised machine learning algorithms.
We construct three feature sets that capture different aspects of the user activity in the diffusion process. 
Our experiments showed that the best-performing feature set accounts for the stance composition of the threads in which a user chooses to engage.
Notably, our activity feature sets outperform a textual baseline that encodes the content that the user emits.

One difficulty we met is the lack of ground truth, i.e., the stance for Reddit users at each time interval. 
To obtain the ground truth, we transferred a model trained on a Twitter dataset. 
However, the underlying distribution of language and structure of the two platforms differ.
The transfer labeling risks introducing inaccuracies, and the performances would probably be better if the Reddit users' correct labels were available. This is a perspective of this work.

\section*{Acknowledgement}
This work was partially supported by IDEXLYON ACADEMICS Project ANR-16-IDEX-0005 of the French National Research Agency, Facebook Research under the Content Policy Research Initiative grants, and the Defence Science and Technology Group of the Australian Department of Defence.
We thank Keneth Benoit, who generously shared the Twitter dataset of Brexit discussions~\citep{benoit_network_nodate}.

 \bibliographystyle{splncs04nat}

\begin{thebibliography}{29}
\providecommand{\natexlab}[1]{#1}
\providecommand{\url}[1]{\texttt{#1}}
\providecommand{\urlprefix}{URL }
\expandafter\ifx\csname urlstyle\endcsname\relax
  \providecommand{\doi}[1]{doi:\discretionary{}{}{}#1}\else
  \providecommand{\doi}{doi:\discretionary{}{}{}\begingroup
  \urlstyle{rm}\Url}\fi

\bibitem[{{Amador Diaz Lopez} et~al.(2017){Amador Diaz Lopez},
  Collignon-Delmar, Benoit, and Matsuo}]{benoit_network_nodate}
{Amador Diaz Lopez}, J.C., Collignon-Delmar, S., Benoit, K., Matsuo, A.:
  {Predicting the Brexit Vote by Tracking and Classifying Public Opinion Using
  Twitter Data}. Statistics, Politics and Policy \textbf{8}(1), 85--104 (sep
  2017), ISSN 2194-6299

\bibitem[{Bail et~al.(2018)Bail, Argyle, Brown, Bumpus, Chen, Hunzaker, Lee,
  Mann, Merhout, and Volfovsky}]{Bail2018}
Bail, C.A., Argyle, L.P., Brown, T.W., Bumpus, J.P., Chen, H., Hunzaker,
  M.B.F., Lee, J., Mann, M., Merhout, F., Volfovsky, A.: {Exposure to opposing
  views on social media can increase political polarization.} PNAS
  \textbf{115}(37), 9216--9221 (2018)

\bibitem[{Banisch and Olbrich(2019)}]{Banisch2019}
Banisch, S., Olbrich, E.: {Opinion polarization by learning from social
  feedback}. Journal of Mathematical Sociology \textbf{43}(2), 76--103 (apr
  2019), ISSN 0022-250X

\bibitem[{Barber{\'{a}}(2014)}]{Barbera2014}
Barber{\'{a}}, P.: {How Social Media Reduces Mass Political Polarization.
  Evidence from Germany, Spain, and the U.S.} Midwest Pol. Sci. Assoc. p.~44
  (2014)

\bibitem[{Bonchi et~al.(2019)Bonchi, Gionis, Ordozgoiti, Galimberti, and
  Ruffo}]{Bonchi2019}
Bonchi, F., Gionis, A., Ordozgoiti, B., Galimberti, E., Ruffo, G.: {Discovering
  polarized communities in signed networks}. In: CIKM, pp. 961--970 (2019)

\bibitem[{Chen and Guestrin(2016)}]{chen2016xgboost}
Chen, T., Guestrin, C.: Xgboost: A scalable tree boosting system. In: KDD, pp.
  785--794, ACM (2016)

\bibitem[{Das et~al.(2014)Das, Gollapudi, and Munagala}]{Das2014}
Das, A., Gollapudi, S., Munagala, K.: {Modeling opinion dynamics in social
  networks}. In: WSDM, pp. 403--412 (2014)

\bibitem[{De-Wit et~al.(2019)De-Wit, Brick, and {Van Der Linden}}]{De-Wit2019}
De-Wit, L., Brick, C., {Van Der Linden}, S.: {Are Social Media Driving
  Political Polarization? Battles} (2019)

\bibitem[{Dubois and Blank(2018)}]{Dubois2018}
Dubois, E., Blank, G.: {The echo chamber is overstated: the moderating effect
  of political interest and diverse media}. Inf. Comm. and Soc. \textbf{21}(5),
  729--745 (2018)

\bibitem[{Friedman(2001)}]{friedman2001greedy}
Friedman, J.H.: Greedy function approximation: a gradient boosting machine.
  Annals of statistics pp. 1189--1232 (2001)

\bibitem[{Garimella et~al.(2016)Garimella, {De Francisci Morales}, Gionis, and
  Mathioudakis}]{Garimella2016}
Garimella, K., {De Francisci Morales}, G., Gionis, A., Mathioudakis, M.:
  {Quantifying controversy in social media}. In: WSDM, vol.~1, pp. 33--42 (jan
  2016)

\bibitem[{Garimella et~al.(2017{\natexlab{a}})Garimella, {De Francisci
  Morales}, Gionis, and Mathioudakis}]{Garimella2017}
Garimella, K., {De Francisci Morales}, G., Gionis, A., Mathioudakis, M.:
  {Reducing controversy by connecting opposing views}. In: WSDM, pp. 81--90
  (2017{\natexlab{a}})

\bibitem[{Garimella et~al.(2017{\natexlab{b}})Garimella, Morales, Gionis, and
  Mathioudakis}]{Garimella2017a}
Garimella, K., Morales, G.D.F., Gionis, A., Mathioudakis, M.: {Exposing Twitter
  Users to Contrarian News}  (mar 2017{\natexlab{b}})

\bibitem[{Garimella et~al.(2017{\natexlab{c}})Garimella, Morales, Gionis, and
  Mathioudakis}]{Garimella2017c}
Garimella, K., Morales, G.D.F., Gionis, A., Mathioudakis, M.: {The Ebb and Flow
  of Controversial Debates on Social Media}. In: ICWSM, pp. 524--527
  (2017{\natexlab{c}})

\bibitem[{Gillani et~al.(2018)Gillani, Yuan, Saveski, Vosoughi, and
  Roy}]{Gillani2018}
Gillani, N., Yuan, A., Saveski, M., Vosoughi, S., Roy, D.: {Me, My Echo
  Chamber, and I}. In: WWW, pp. 823--831, ACM (2018)

\bibitem[{Graells-Garrido et~al.(2016)Graells-Garrido, Lalmas, and
  Quercia}]{Graells-Garrido2013}
Graells-Garrido, E., Lalmas, M., Quercia, D.: {Data Portraits: Connecting
  People of Opposing Views}. In: International Conf. on Intelligent User
  Interfaces. (2016)

\bibitem[{Gr{\v{c}}ar et~al.(2017)Gr{\v{c}}ar, Cherepnalkoski, Mozeti{\v{c}},
  and {Kralj Novak}}]{Grcar2017}
Gr{\v{c}}ar, M., Cherepnalkoski, D., Mozeti{\v{c}}, I., {Kralj Novak}, P.:
  Stance and influence of twitter users regarding the brexit referendum.
  Comp.Soc.Net. \textbf{4}(1), 1--25 (2017)

\bibitem[{Howard and Kollanyi(2016)}]{howard2016bots}
Howard, P.N., Kollanyi, B.: {Bots, \#StrongerIn, and \#Brexit: computational
  propaganda during the UK-EU referendum}. Available at SSRN 2798311  (2016)

\bibitem[{Hughes and Palen(2009)}]{hughes2009twitter}
Hughes, A.L., Palen, L.: Twitter adoption and use in mass convergence and
  emergency events. Int. jour. of emergency management \textbf{6}(3-4),
  248--260 (2009)

\bibitem[{Kim et~al.(2019)Kim, Graham, Wan, and Rizoiu}]{Kim2019}
Kim, D., Graham, T., Wan, Z., Rizoiu, M.A.: {Analysing user identity via
  time-sensitive semantic edit distance (t-SED): a case study of Russian trolls
  on Twitter}. Journal of Computational Social Science \textbf{2}(2), 331--351
  (jul 2019)

\bibitem[{Liao and Fu(2013)}]{Liao2013}
Liao, Q.V., Fu, W.T.: {Beyond the filter bubble: Interactive effects of
  perceived threat and topic involvement on selective exposure to information}.
  In: Human Factors in Computing Systems, pp. 2359--2368 (2013)

\bibitem[{Matakos et~al.(2017)Matakos, Terzi, and Tsaparas}]{Matakos2017}
Matakos, A., Terzi, E., Tsaparas, P.: {Measuring and moderating opinion
  polarization in social networks}. DAMI \textbf{31}(5), 1480--1505 (2017)

\bibitem[{Messing and Westwood(2014)}]{Messing2014}
Messing, S., Westwood, S.J.: {Selective Exposure in the Age of Social Media}.
  Communication Research \textbf{41}(8), 1042--1063 (dec 2014), ISSN 0093-6502

\bibitem[{Mishra et~al.(2018)Mishra, Rizoiu, and Xie}]{Mishra2018}
Mishra, S., Rizoiu, M.A., Xie, L.: {Modeling Popularity in Asynchronous Social
  Media Streams with Recurrent Neural Networks}. In: ICWSM, pp. 1--10 (2018)

\bibitem[{Musco et~al.(2018)Musco, Musco, and Tsourakakis}]{Musco2018}
Musco, C., Musco, C., Tsourakakis, C.E.: {Minimizing Polarization and
  Disagreement in Social Networks}. In: WWW, pp. 369--378 (2018)

\bibitem[{Pushshift(2019)}]{pushshift}
Pushshift: Pushshift. \url{https://pushshift.io/} (2019)

\bibitem[{Rama et~al.(2017)Rama, Garimella, and Weber}]{Rama2017}
Rama, V., Garimella, K., Weber, I.: {A Long-Term Analysis of Polarization on
  Twitter}. In: ICWSM, pp. 528--531 (2017)

\bibitem[{Rizoiu et~al.(2018)Rizoiu, Graham, Zhang, Zhang, Ackland, and
  Xie}]{Rizoiu2018a}
Rizoiu, M.A., Graham, T., Zhang, R., Zhang, Y., Ackland, R., Xie, L.:
  {{\#}DebateNight: The Role and Influence of Socialbots on Twitter During the
  1st 2016 U.S. Presidential Debate}. In: ICWSM, pp. 1--10 (2018)

\bibitem[{Sch{\"u}tze et~al.(2008)Sch{\"u}tze, Manning, and
  Raghavan}]{schutze2008introduction}
Sch{\"u}tze, H., Manning, C.D., Raghavan, P.: Introduction to information
  retrieval, vol.~39. Cambridge University Press Cambridge (2008)

\end{thebibliography}

\clearpage
\appendix
\section{Appendix}

This document is accompanying the submission \textit{Linking User Opinion Dynamics and Online Discussions}.
The information in this document complements the submission, and it is presented here for completeness reasons.
It is not required for understanding the main paper nor for reproducing the results.

\subsection{Posting analysis and dataset profiling}

\begin{figure}[htbp]
	\centering
	\newcommand\myheightA{0.18} \subfloat[]{
		\includegraphics[height = \myheightA\textheight]{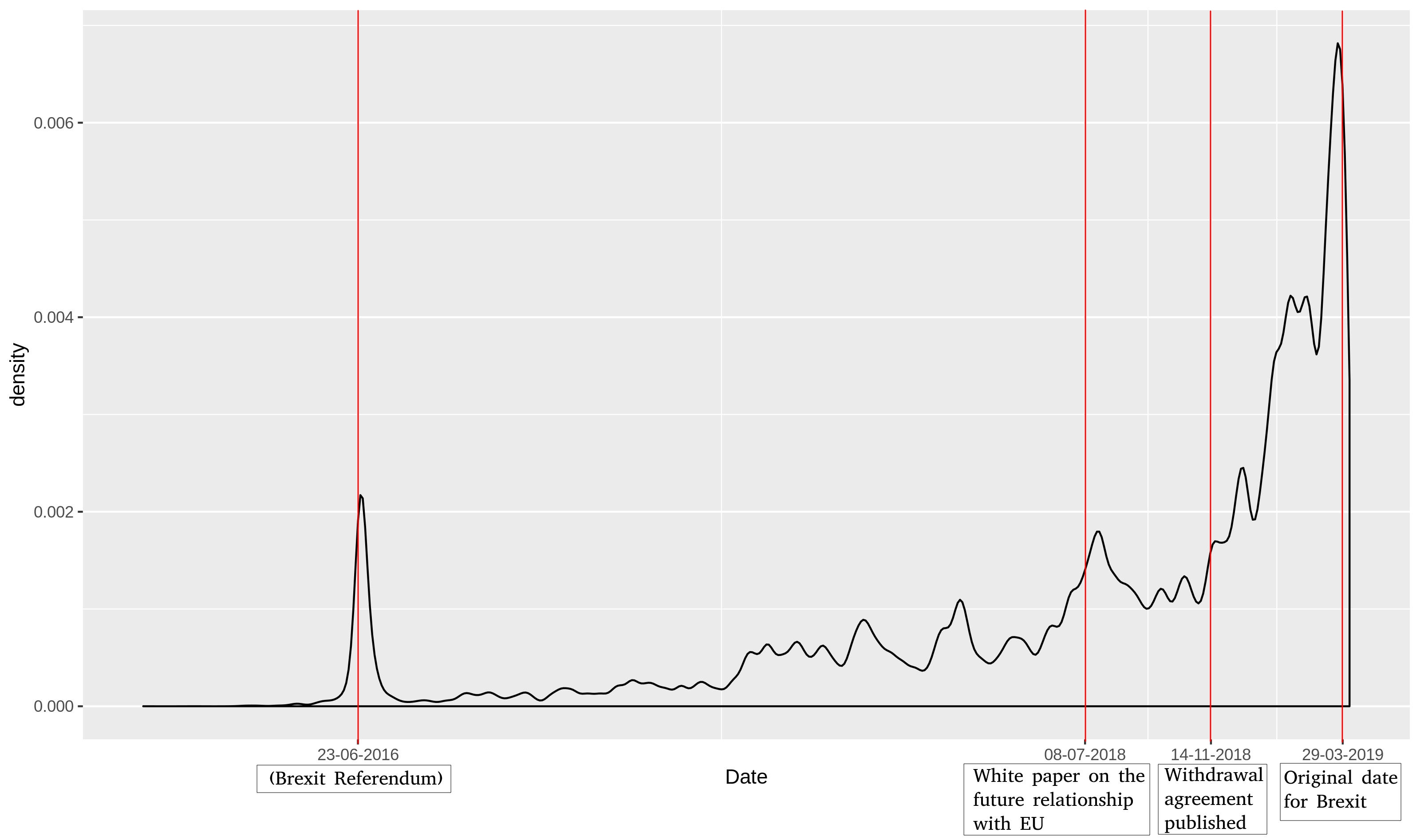}
		\label{fig:figure_4}
	}\subfloat[]{
		\includegraphics[height = \myheightA\textheight]{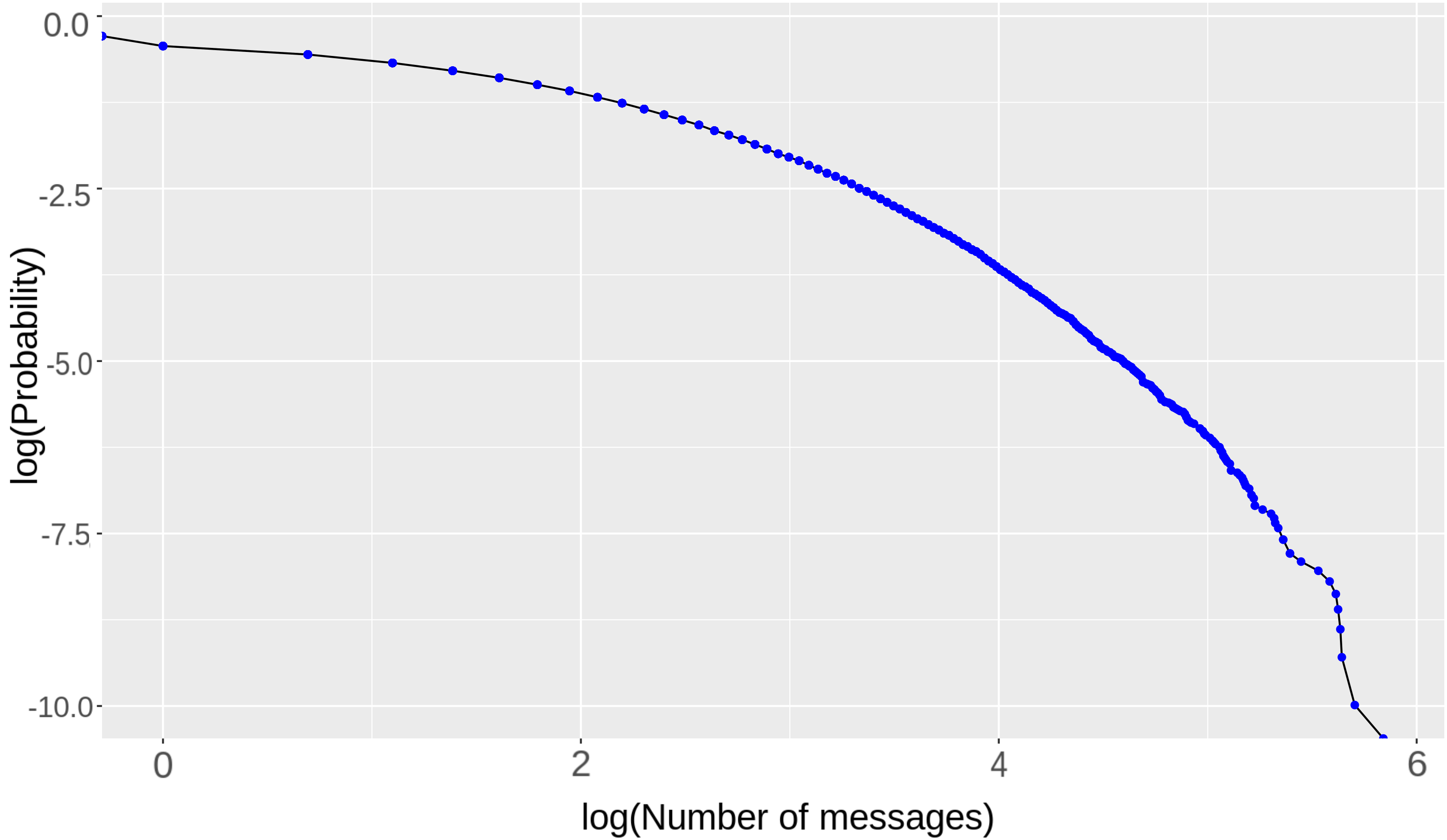}\label{fig:figure_6}
	}\caption{
		\textbf{(a)} Time distribution of the submissions collected from Reddit (subreddit brexit), between November, 2015 and April, 2019. 
		\textbf{(b)} Complementary Cumulative Distribution Function of the number of messages sent by each user.
	}

\end{figure}

\textbf{Distribution of Reddit posts over time.} 
The time distribution of the collected messages is presented in \cref{fig:figure_4}. 
We see that generally, there is an ascending trend, which may indicate the growing importance of this subject on the online social platform. 
We observe a spike for June 2016 because, on the 13th of June 2016, British people voted in the national referendum. 
On the other hand, most messages were submitted messages in February - March 2019. 
This is a consequence of the Brexit process's official schedule, which should have completed in March 2019.

The vast majority of posted messages are comments ($91 \%$) and only $9 \%$ are initial posts starting threads.
Moreover, $20\%$ of all the unique authors are thread initiators, meaning they only send a single message, starting a discussion thread, in which they never post again. 
On the other hand, $19\%$ of the authors are both thread starters and commenters, meaning that they start threads and participate actively in the discussions, posting answers in their own started thread or getting involved in other discussions. 
The majority of the unique users ($61 \%$) are commenters only; they never start discussions but usually engage in them.

\Cref{fig:figure_6} presents the Complementary Cumulative Distribution Function of the number of messages per user (in log-log scale). 
It shows that most users send only a few messages, whereas a few users send most messages in the observed interval. 
This is a known phenomenon in online social activity~\cite{Mishra2018}, where a minority of users produce and exchange the vast majority of content while the rest are silent consumers.

\subsection{Additional graphics and information}

\begin{figure}[htbp]
	\centering
	\newcommand\mywidth{0.49} \subfloat[]{
		\includegraphics[width = \mywidth\textwidth,page=1]{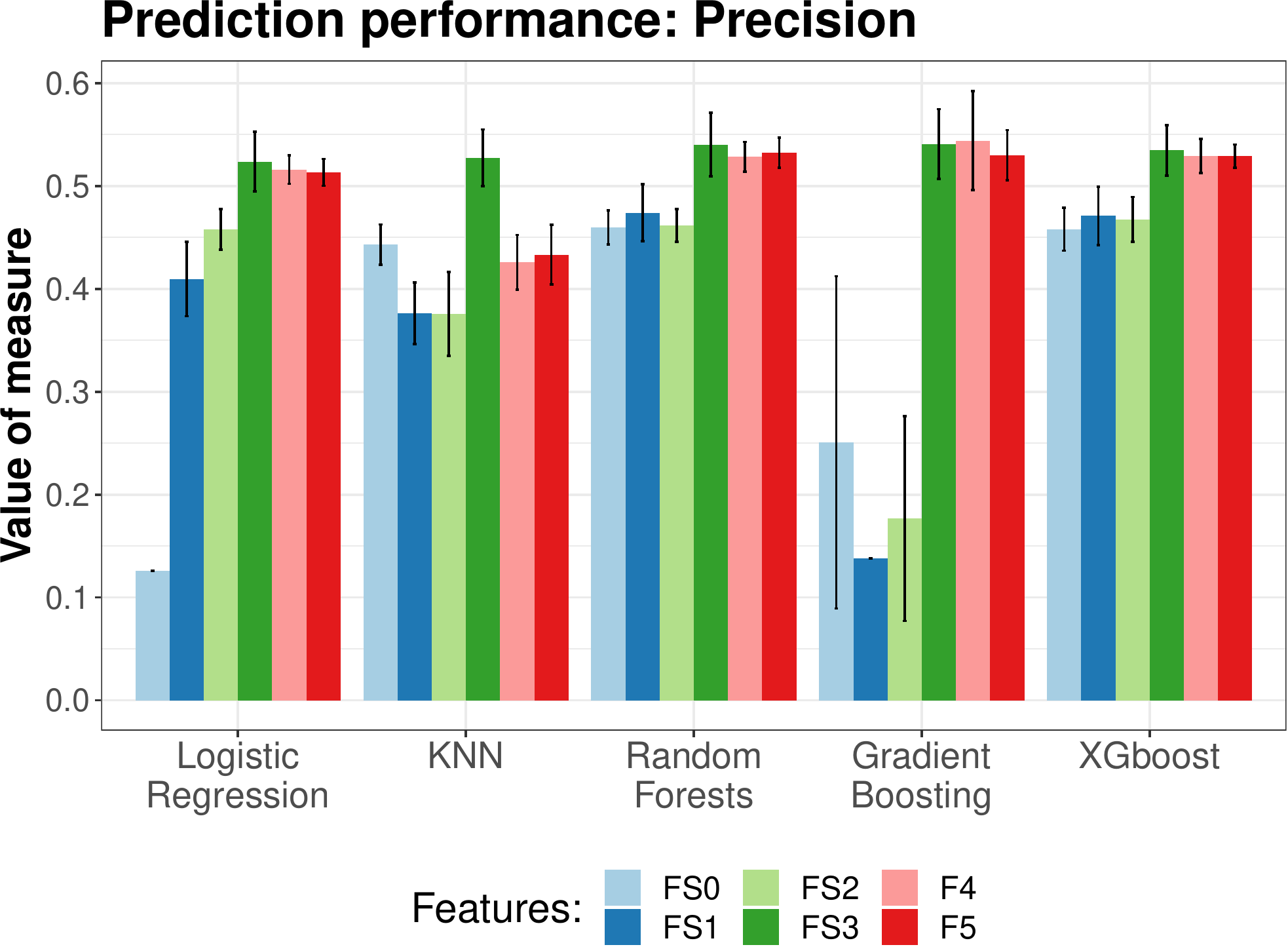}
		\label{fig:precision}
	}\subfloat[]{
		\includegraphics[width = \mywidth\textwidth,page=2]{All-models_barplots}\label{fig:recall}
	}\caption{
		\textbf{Additional performance measures for predicting future stance with different feature sets:}
		macro-precision \textbf{(a)} and macro-recall \textbf{(b)}.
	}

\end{figure}

\Cref{tab:time-interval-split}, \cref{fig:precision,fig:recall} show the additional information promised in the main text.

\begin{table}[htp]
	\scriptsize
	\newcommand\mywidth{8.5cm} \centering
	\caption{
		The real-life events used to split the Reddit dataset into timeframes, together with basic profiling of the timeframes.
	}
	\label{tab:time-interval-split}

	\begin{tabular}{clrrp{\mywidth}}
		\toprule
		\textbf{Int.} & \multicolumn{1}{c}{\textbf{Start date}} & \multicolumn{1}{c}{\textbf{Posts}} & \multicolumn{1}{c}{\textbf{Users}} & \multicolumn{1}{c}{\textbf{Important events in the Brexit chronology}} \\ \midrule
		T1 & 2015-11-16 & 3367 & 1268 & \makecell[t{p{\mywidth}}]{
			\textbf{23 June 2016}\\
			The UK holds a referendum on whether to leave the European Union (EU). 51.9\% of voters vote to leave.\\
			\textbf{24 June 2016}\\
			David Cameron announces his resignation as Prime Minister.} \\ 
			
		T2 & 2016-06-25 & 6265 & 1623 & \makecell[t{p{\mywidth}}]{
			\textbf{13 July 2016}\\
			Theresa May accepts the Queen's invitation to form a government} \\
			
		T3 & 2016-07-14 & 3084 & 810 & \makecell[t{p{\mywidth}}]{
			\textbf{27 July 2016}\\
			The European Commission nominates French politician Michel Barnier as European Chief Negotiator for the United Kingdom Exiting the European Union.\\
			\textbf{7 December 2016}\\
			The UK House of Commons votes 461 to 89 in favour of May’s plan to trigger Article 50 by the end of March 2017} \\

		T4 & 2016-12-08 & 1466 & 320 & \makecell[t{p{\mywidth}}]{
			\textbf{24 January 2017}\\
			UK Supreme Court rules that Parl. must pass legislation to authorize the trigger of Art. 50.\\
			\textbf{26 January 2017}\\
			The UK Gov. introduces a 137-word bill in Parl. to empower May to initiate Brexit by triggering Art 50.} \\

		T5 & 2017-01-27 & 2300 & 431 & \makecell[t{p{\mywidth}}]{
			\textbf{16 March 2017}
			The bill receives Royal Assent.\\
			\textbf{29 March 2017}\\
			A letter from May is handed to President of the European Council Donald Tusk to invoke Article 50, starting a two year process with the UK due to leave the EU on 29 March 2019.} \\

		T6 & 2017-03-30 & 4102 & 557 & \makecell[t{p{\mywidth}}]{
			\textbf{19 June 2017}
			Brexit negotiations commence.} \\

		T7 & 2017-06-20 & 54505 & 2339 & \makecell[t{p{\mywidth}}]{
			\textbf{6 July 2018}\\
			A UK White paper on The future relationship between the United Kingdom and the European Union is finalized.\\
			\textbf{8 July 2018}\\
			Davis resigns as Secretary of State for Exiting the EU. Dominic Raab is appointed as his successor the following day.} \\

		T8 & 2018-07-09 & 23067 & 1479 & \makecell[t{p{\mywidth}}]{
			\textbf{21 September 2018}
			EU rejects the UK white paper.} \\

		T9 & 2018-09-22 & 15385 & 1480 & \makecell[t{p{\mywidth}}]{
			\textbf{14 November 2018}
			Brexit withdrawal agreement published.\\
			\textbf{15 November 2018}\\
			Raab resigns as Secretary of State for Exiting the EU. Stephen Barclay is appointed as his successor the following day.} \\

		T10 & 2018-11-16 & 3718 & 732 & \makecell[t{p{\mywidth}}]{
			\textbf{25 November 2018}\\
			Other 27 EU Member States endorse the Withdrawal Agreement.} \\

		T11 & 2018-11-26 & 25468 & 2485 & \makecell[t{p{\mywidth}}]{
			\textbf{15 January 2019}\\
			First meaningful vote held on the Withdrawal Agreement in the UK House of Commons. The UK Gov. is defeated by 432 votes to 202} \\

		T12 & 2019-01-16 & 54850 & 4489 & \makecell[t{p{\mywidth}}]{
			\textbf{12 March 2019}\\
			Second meaningful vote on the Withdrawal Agreement with the UK Government defeated again by 391 votes to 242.\\
			\textbf{14 March 2019}\\
			UK Gov. motion passes 412 to 202 to extend the Article 50 period.} \\

		T13 & 2019-03-15 & 9119 & 1836 & \makecell[t{p{\mywidth}}]{
			\textbf{20 March 2019}\\
			May requests the EU extend the Article 50 period until 30 June 2019.\\
			\textbf{21 March 2019}\\
			The European Council offers to extend the Article 50 period until 22 May 2019 if the Withdrawal Agreement is passed by 29 March 2019 but, if it does not, then the UK has until 12 April 2019 to indicate a way forward. The extension is formally agreed the following day.} \\
	
		T14 & 2019-03-22 & 13414 & 2444 & \makecell[t{p{\mywidth}}]{
			\textbf{29 March 2019}\\
			The original end of the Article 50 period and the original planned date for Brexit. Third vote on the Withdrawal Agreement after being separated from the Political Declaration. UK Government defeated again by 344 votes to 286.} \\
	
		T15 & 2019-03-30 & 9509 & 1840 & \makecell[t{p{\mywidth}}]{
			\textbf{5 April 2019}\\
			May requests for a second time that the EU extend the Article 50 period until 30 June 2019.} \\

		& 2019-04-05 &  &  & \textit{--dataset end--} \\ \bottomrule
	\end{tabular}

\end{table} 
\end{document}